# Spatial and Temporal Fluctuations of an Active Brownian Particle in an Optical Trap

Chong Shen[1], Lanfang Li[2], Zhiyu Jiang[1], H. D. Ou-Yang[1,2]
1. Department of Physics, Lehigh University, 16 Memorial Dr. E, Bethlehem, PA 18015, USA
2. Emulsion Polymers Institute, Lehigh University, 16 Memorial Dr. E, Bethlehem. PA 18015, USA

## Abstract

A colloidal suspension of active Brownian particles (ABPs) driven by controllable forces into directed or persistent motions can serve as a model for understanding the biological systems. Experiments and numerical simulations are established to investigate the motions of an ABP, a single, induced-charge electrophoretic (ICEP) metallic Janus particle, confined in a quadratic potential well. On the one hand, 1-D position histograms of the trapped active particle, behaving differently from that of a Boltzmann distribution, reveal a splitting from a single peak of the ABP's positional distribution to a bimodal distribution. Decoupling the thermal and non-thermal contributions from the overall histogram is non-trivial. However, the two contributions can be examined by convoluting numerically generated thermal and non-thermal contributions into a full histogram. On the other hand, temporal fluctuations analyzed by the power spectral density (PSD), reveal two unique frequencies characterizing the stiffness of the trap and the rotational diffusion of the particle, respectively. Connections between the spatial and temporal fluctuations are obtained by the separate analysis of the temporal and spatial fluctuations of an ABP trapped in a quadratic potential well. This study reveals how thermal and nonthermal fluctuations play against each other in a confined environment

## 1. Introduction

Active matter, which is composed of the active agents, is characterized by the ability to transduce energy to the movement. [1] Active matter is important as the biological system can be categorized as an active system. In active matter, non-thermal fluctuations play an important role. [1]

The non-thermal fluctuations have been studied by calculated the fluctuation power spectral (PSD) in active gel [2], living cell [3], and red blood cell [4]. These work observed a raise of PSD value at low frequency. However, more characteristic of the active agent could be obtained at an even lower frequency [5]. Also, Fodor et al [6] suggest using the integration of the PSD to yield the total energy dissipation of the system, which could be questionable because of the fluctuation-dissipation relation didn't hold in the non-equilibrium system. [7]

In order to describe the behavior of a living system using non-equilibrium thermodynamics [8,9], Active Brownian particle system has been used as a model system, in which the particles move at a constant speed with the direction being determined by the rotational diffusion. [10] Previously, Boltzmann distribution was used to determine the spatial distribution of Active Brownian Particle (ABP) under a uniform force field at an effective temperature higher than ambient temperature, which is the classical equilibrium temperature. [11,12] However, the effective temperature treatment still calls for Boltzmann distribution, but recent research studies of ABP system in confinement show the collection of particles near the wall and thus the distribution is not Boltzmann any more. [13–15]. The histogram of particle position (HPP)describes the particle's time-averaged probability at certain position. The simpler case of only considering the active component of the movement has been treated with mean-field theory, however, it did not encompass the thermal movement component. [16] Takatori et. al. performed a set of experiments with chemical phoretic Janus particles in an acoustic trap and revealed a similar behavior to mean-field theory and simulation. [17] But Arjun et al conducted the experiments of a bacteria kicking ABP in a small optical trap showed deviation from the mean-field theory without thermal fluctuation. [18,19] The weak trap strength of the acoustic trap study and the limited dynamic range of the bacterial bath prohibited exploration of a more complete parameter space. Until now, when and how the Boltzmann statistics would fail for ABP is still not clear. [14]

In this work, we use a model system with a large dynamic range to investigate an active particle in a quadratic optical trap, considering both thermal and non-thermal noise. We refer to this particle as active Brownian particle (ABP), in that the rotation of the particle is random as a Brownian particle but the translational motion is actively driven. We analyze the system from both the histogram of particle position (HPP) and fluctuation power spectral density (PSD) aspect. We explain the conditions for Gaussian and non-Gaussian particle distribution, as well as the conditions where an effective temperature is a good parameter to use in HPP. We quantified the shape and characteristic frequency in PSD, which related to the particle speed and rotational relaxation. We also uncover the relationship between HPP and PSD and discussing the definition and relationship of effective temperature in active matter system. We use effective temperatures from different definition to quantified the dissipation under different trap stiffness.

## 2. Materials and Method:

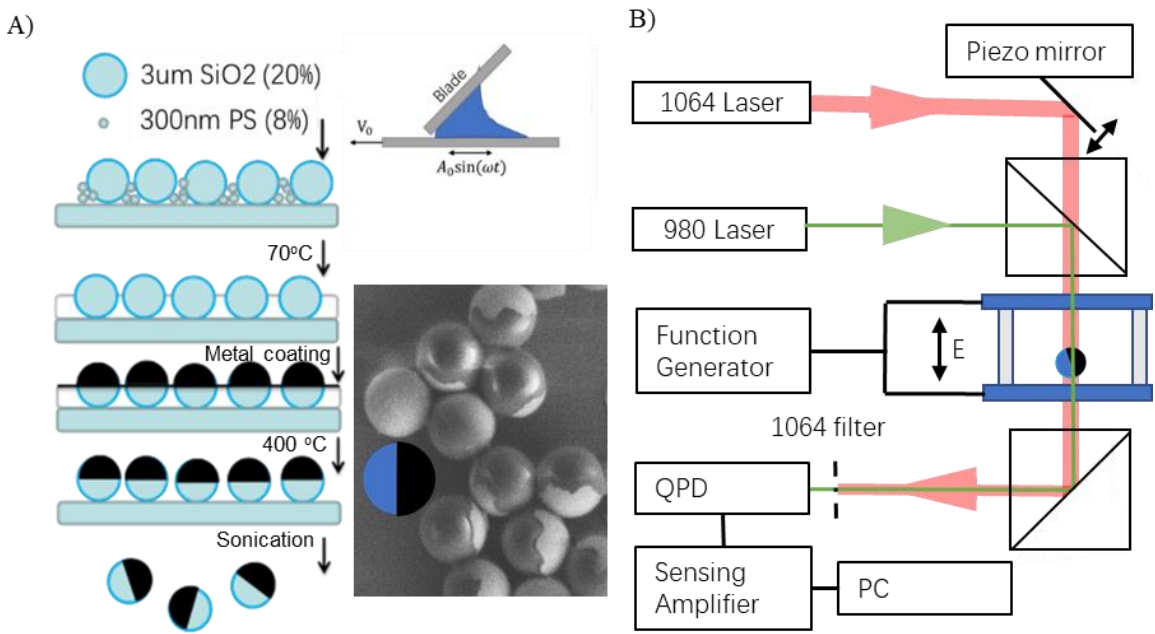

Fig.1 Experimental setup of active Brownian particle in an optical trap. A) The fabrication process of metallic dielectric Janus particles. Insert: SEM image of experimental Janus particles: 3.0 um Silica beads partially coated with iridium (Ir) with a sketch of the Janus particle. B) Schematic of optical trap set-up. the co-aligned optical trapping (1064nm laser) and tracking beam (980 nm laser) provide simultaneous trapping and particle position measurement with high spatial and temporal resolution. The Janus particles are loaded in ITO coated glass cell capable of applying an electric field perpendicular to the substrate.

## 2.1 Fabrication of Janus particles

We synthesized 3um metallic Janus particles by depositing a thin film of metal on 3um silica particles (as purchased, Bangs Laboratory, SS05N). [20] First, the 3 um particles (dry powder) were mixed with 300 nm polystyrene (PS) latex (40%wt) suspension at 20% volume particle 8% volume PS in water. The mixture was then deposited by vibration-assisted convective deposition method [21] to create a monolayer of 3um base silica particles on a glass substrate. After drying under ambient condition, the composite microsphere monolayer on glass was then heated at 240 °C for a few minutes so the PS particles melted into a 1650 nm PS film covering the bottom half of the $SiO_2$ spheres. The monolayer of $SiO_2$ spheres, with bottom half immersed in a PS layer, was then loaded into a sputter coater (Poloran E5100) to receive a 30nm coating of Ir on the top half surface. The sample was then heated at 500 °C (Lindberg/Blue M™ Moldatherm™ Box Furnaces, BF51748C-1) for 1 hour to remove the polystyrene. The sample was then placed in a water filled centrifuge tube and sonicated in a sonication bath (Branson, 1510) for 6 hours to release the particles, forming hemispherically coated Janus particles suspension in water. The SEM (Hitachi 4300SE ) micrograph of the resultant Janus particles is shown in Fig.1 A

## 2.2 Drive a Janus particle in E field by ICEP

Janus particle can be driven by a uniform AC electric field to move in a direction perpendicular to the field. [22] We used ITO coated glass slides as transparent electrodes. The model system contained a suspension of Janus particles sandwiched between two ITO coated glass slides with 50um separation, showing in Fig.1 B). A 5kHz AC voltage was applied across the electrodes to create a uniform electric field perpendicular to the electrodes. This electric field can drive the Janus particles in the horizontal direction by a mechanism of induced-charge electrophoresis (ICEP), making it a model two dimensional (2D) ABP system. [19] We tracked the particle to find the trajectory from video by ImageJ with Mosaic Suit. [23]

## 2.3 Optica trapping of a Janus particle and concurrent tracking of particle position

We used a 1064nm wavelength laser to create an optical trap, as shown in Fig 1.C). The stiffness of the trap was measured to be 3.0pN/um with 25mw laser power by passive and active microecology. [24,25] The trap is approximately a quadratic potential well. To avoid large scattering force on the metal cap of the particle, which has been shown to make optical trapping unstable, [26,27] we use particles with metal cap size areas smaller than a hemisphere to reduce the scattering effect. By comparing a Janus particle to that of uncoated particle motion, they are indistinguishable, indicating that the light scattering in these particles is negligible.

To achieve a high temporal and spatial detection resolution of the particle position, a 980nm wavelength laser was applied to serve as a tracking beam that was aligned collinearly with the trapping laser beam. A Quadrant PhotoDiode (in-house design) was at the end of the tracking beam, analyzing the deflection of the tracking beam. As the particle information is carried by the deflection of the tracking beam, the particle position can be measured from the position of the deflected tracking beam. Such detection scheme can achieve 5nm spatial resolution at 1 kHz sampling rate.

## 2.4 Numerical simulation method

To understand the motion of the particle, a first principle calculation was performed using the generalized Langevin equation (Eq 1). [19,28]

$$m\ddot{r} = \hat{\theta}F_0 - \xi\dot{r} - k_{OT}r + \sqrt{2k_BT}W_{noise} \qquad (1)$$

In overdamped condition, the equation becomes Eq 2,

$$0 = \hat{\theta}v_0D - \xi\dot{r} - k_{OT}r + \sqrt{2k_BT}W_{noise} \qquad (2)$$

where $v_0$, $k_{OT}$ can be experimentally determined, the drag-coefficient being self-drag, Wnoise from a random number generator matching a Gaussian profile. With no fitting parameters, the simulation captures the essence of the motion of the particle at a first principle level.

## 3. Results and discussion

### 3.1 A Single Particle in Free Space

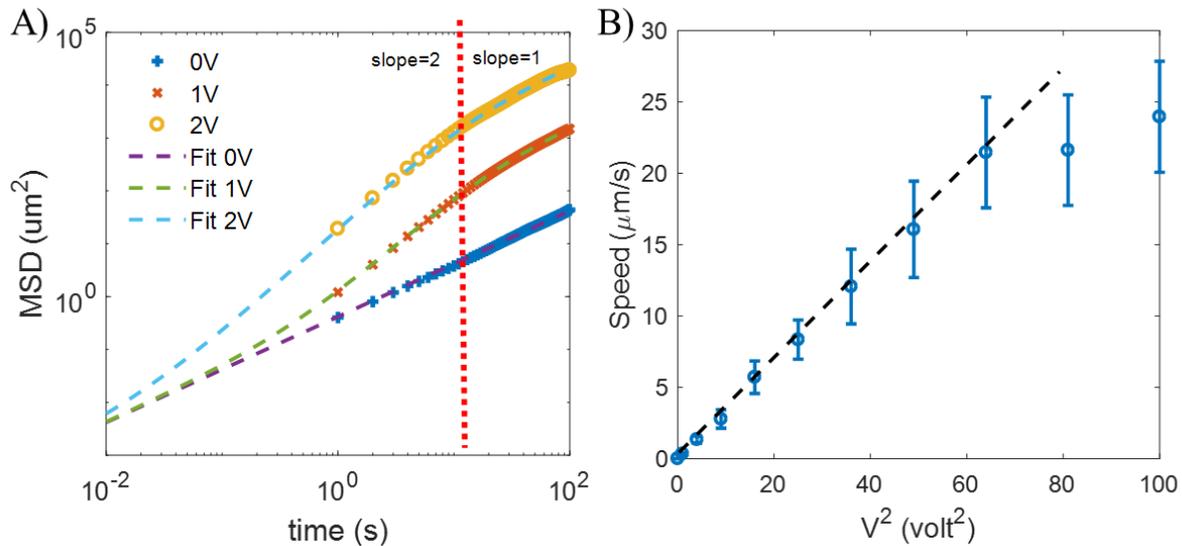

Fig. 2 Characterization of particle movement in an optical trap. A): MSD data from experiment under different voltages. B) Swim speed under different voltages. At lower voltages the particle swim speed show linear relationship with applied voltage squared.

Using video analysis of particles in the experimental cell without any confinement, we can track the position of the particles, and calculate mean square displacement (MSD) for any given particle in the field of view. The experimental MSD vs time for particles at varying applied voltage is shown in Fig 2A. At zero applied voltage, the particle shows characteristics of Brownian motion, exhibiting slope of 1 in the MSD vs time graph. Using $D_t = \frac{dMSD}{4dt}$ and measured MSD curve, it can be determined that the zero voltage diffusivity is $0.053 \pm 0.008 um^2/s$. The Janus particles in this experiment are heavy enough to stay relatively close to the glass substrate typically do not stick to the substrate. From electric conductivity measurement, it was estimated that the Debye length for the particles is ~ 12nm. Using Fasen's law for a single wall, the distance between the particle and ITO coated glass substrate is ~ 12 nm. From the Stokes-Einstein equation $D_t = \frac{k_B T}{6\pi \eta r}$, the translational diffusivity is 0.145 um²/s at a distance far away from the substrate, so there is some substrate effect, we believe this effect is similar across all samples regardless of applied voltage. ...

When a non-zero AC electric field is applied, the Janus particles can undergo induced charge dielectrophoresis (ICEP), as predicted theoretically by Squires and Bazant, [29] and subsequently observed experimentally by Gangwal et al. [22] The metallic-dielectric Janus particle will move by ICEP away from the metal cap at a velocity,

$$v(ICEP) = \frac{9}{64} \frac{\epsilon a}{\eta(1+\delta)} E^2 \qquad (3)$$

where $\epsilon$ is the electric permittivity, η is the viscosity of the bulk solvent, $a$ is the radius of the particle, E is the electric field, and δ is the ratio of the capacitances of the compact and diffuse layers. With a uniform thickness experimental cell used in the experiment, the ICEP velocity should be proportional to the square of the voltage.

We use the ICEP velocity as the active swim mechanism. With an applied voltage, there are two slopes in the MSD vs time curve: at the short-time a slope of 2 which signifies a ballistic motion; at the long-time a slope of 1 which is consistent with Brownian diffusion motion but with a higher diffusivity. The transition between the mode of motion can be characterized by relaxation time [30] for ABP.

$$MSD = 4D_t t + 2v_0^2 \tau_r [t - \tau_r(1 - e^{-t/\tau_r})] \qquad (4)$$

Where $v_0$ active swim speed, $\tau_r$ is rotational relaxation time. From our experimental data, at varying applied AC field, the rotational relaxation time stays constant. Fitting the experimental MSD data yields $\tau_r = 17.8 \pm 1.4/s$. Which is close to relaxation time calculated for particles with a radius of 1.5 microns using Stokes-Einstein relation: $1/D_r = \frac{8\pi \eta r^3}{k_B T} = 20.6 s$. Plotting active speed $v_0$ and applied voltage squared, we see a linear correlation between the two at voltage up to 20 um/s, as shown in Fig 2B.

The MSD vs time curve in Fig 2A can be fit very well with equation 2, this gives us more insight into the motion of a particle without confinement, and one can use MSD to determine whether there is active speed from the shape of MSD curve. Without the active swim speed driven by an electric field, The particle has a translational and rotational motion like a Brownian particle. When the active swim speed is induced by electric field, at very short time scale, the motion is still thermal diffusion dominated, with a slope of 1 in the MSD vs time curve. A ballistic motion appears at a time scale of $\tau_d$= $D_t/v_0^2$, where $D_t$ is translational diffusivity, the MSD vs time curve will start to exhibit a slope of 2, and the onset time $\tau_a$ is inversely proportional to square of active swim speed. At long time scale of the MSD vs time curve appears to be similar to thermal diffusion again with a slope of 1 but with a higher diffusivity, the transition into this phase is at a characteristic time scale $\tau_r = 1/Dr$, the rotational relaxation time, beyond which rotational diffusion dominates. In our system, $\tau_d < \tau_r$, but it is possible that in some active system, $\tau_d < \tau_r$, then there will be a slope of 1 throughout, one will not be able to conclude that the particle is not active just with a slope of 1 in the MSD vs time graph.

## 3.2 Single Particle in an Optical Trap

Non-equilibrium fluctuation plays an important role in serval biological functions. To study the non-equilibrium fluctuation, we use a simple model system, which is a single active particle in optical trap confinement, as an optical trap is almost ideally quadratic, strong, does not impose a hard boundary, and resembles a lot of biological conditions and process. Optical trap, on the other hand, can not very big, demanding high-resolution time and spatial resolution.

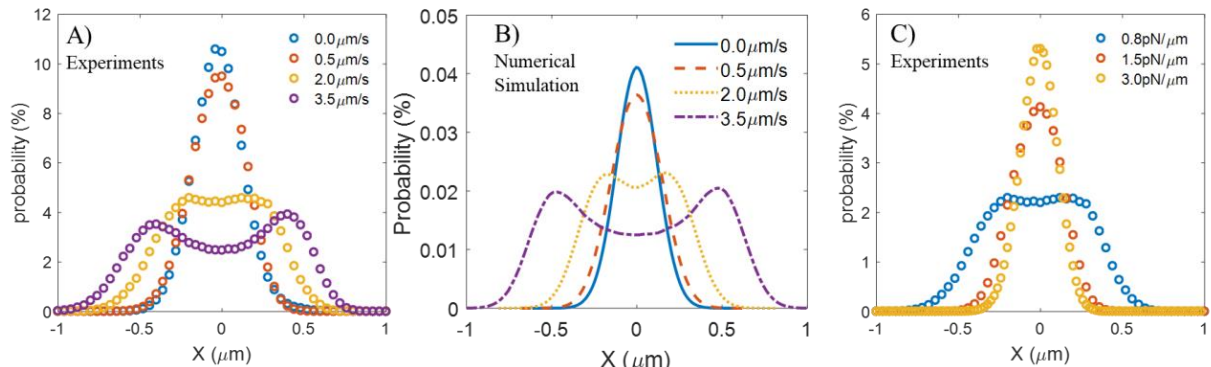

FIG. 3. A) Experimental Histogram of the particle position (HPP) of a 3um Janus particle in a quadratic potential with a spring constant of 1.5pN/um for various 5kHz AC electric field. B) Experimental HPP of a 3um Janus particle under 5V applied voltage ( 2um/s speed), in optical traps with various trap stiffness, the spring constant varying from 0.8 pN/um to 3.0pN/um.C) Numerical simulation of HPPs of a particle with active speed of xxx in a trap with varying trap spring constant.

To analyze the motion of a particle in an optical trap, we used a QPD device to track particle motion. Data analysis is first carried out by employing the Histogram of Particle Position (HPP), defined as the normalized probability of the particle at a particular position in 1-D. [18] The HPP of a Janus particle is shown in the Fig.3 A. At zero voltage, the particle exhibit Gaussian distribution spatially, as seen in Fig 3A data represented by blue circles, indicating Brownian motion. With increasing ICEP electric field, the shapes of HPP change from Gaussian distribution to a double-peak distribution, as seen in Fig 3A. It is well known that the thermal fluctuations follow Gaussian distribution, so it is only possible that the non-Gaussian component is from the active motion. While keeping the spring constant of the optical trap at 1.5pN/um, the increasing ICEP electric field caused the shape of HPP to change from Gaussian distribution to a double-peak distribution, as seen in Fig 3A. The higher the electric field, ie., the active motion speed, the further away the two peaks are from the center and closer to boundary. The higher probability of particles distribution at trap boundary is in agreement with Maggi et al. [31] for active agent collecting at boundary. As far as we know, our experiment was the first to see such boundary collection of active agent in an optical trap, and in such a small scale. While keeping the active speed of the particle constant at 2 um/s, the decreasing optical trap spring constant of the trap from 3pN/um to 0.8pN/um caused the shape of the histograms to change from Gaussian to a flat-top distribution, as seen in Fig 3 B. The bimodal distribution can be understood from the slowing down of the active particle in the trap at a position where the trapping force and active force balance out, which is near the confinement boundary rather than at the center.

To understand the motion of the particle, a first principle simulation was performed using the generalized Langevin equation (Eq 1). [19,28], and its simplified form in overdamped condition Eq 2. With no fitting parameters, the simulation captures the essence of the motion of the particle at a first principle level, as shown in Fig 3 C and D. The active motion component is responsible for the separation fo the Gaussian peak into bimodal double-peak distribution.

We further used the simulation algorithm to decipher the contribution from each component. The spatial distribution of ABP in a trap with thermal noise can be considered as from the active noise and passive noise separately and then combined using convolution. Starting from Langevine equation Eq 1 and Eq 2, then we can split the Langevin equation Eq2 into two Langevin equations with only active motion and only thermal noise.

$$0 = \hat{\theta} F_0 - \xi \dot{r}_1 - k_{OT} r_1 \tag{5}$$

$$0 = -\xi \dot{r}_2 - k_{OT} r_2 + \sqrt{2k_B T} W_{noise} \tag{6}$$

Where $r_1 + r_2 = r$. The first equation describes a pure active motion. The second equation describes thermal motion. Let the histogram from the active motion to be $P_A(r_1)$ and the histogram from the thermal motion to be $P_P(r_2)$. The histogram r can be calculated by:

$$P(r) = P_A * P_P(r) = \int P_A(r_1) P_P(r - r_1) dr_1 \tag{7}$$

The shape of $P(r)$ is determined by $P_A$ or $P_P$ with the large mean squared displacement (MSD).

The results of the convolution method are shown in Fig 4. Under a few different parameter combinations, the yellow curves in Fig 4 are from the original Langevine equation Eq 2, blue curves are with only Brownian motion component using Eq 6, and orange curves are with only active motion and using Eq 5. The purple dash lines are from data using Eq 12, the convolution method. It can be seen that the convolution method yields a perfect fit with the original Langevin equation.

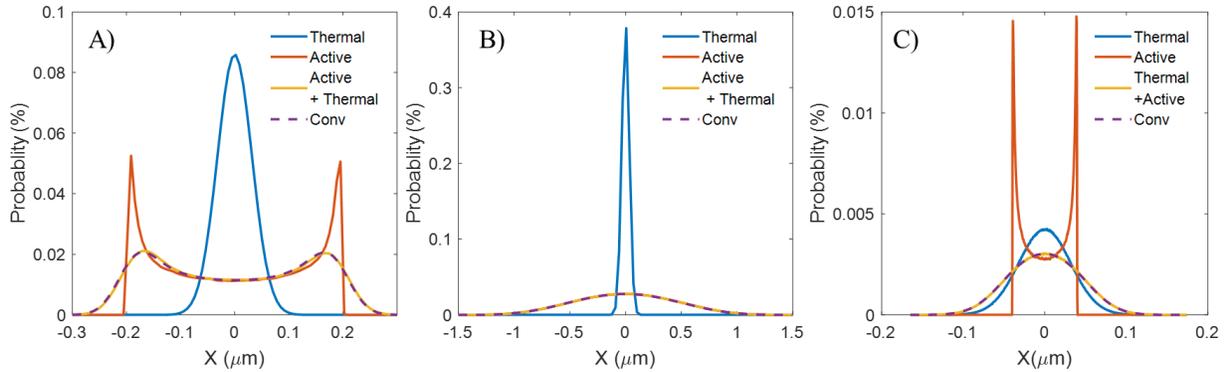

Figure 4. Numerically simulated histogram of particle position (HPP). Thermal noise (solid blue line), active noise (orange solid line), active +thermal noise (yellow solid line) and the convolution (purple dash line) between the histogram of thermal noise and histogram of active noise. Different dynamic parameters: active speed v, and rotational diffusion coefficient $D_r$, are shown. A) v=10um/s $D_r$=3/s  B) v=100um/s $D_r$=500/s  C) v=2um/s $D_r$=3/s

Here, we introduce three characteristic lengths:i) $l_1 = v_0 \eta / k_{OT}$ the equilibrium position in the trap that balances between active force (through particle speed) and trap force ($F = l_1 k_{OT}$), ii) $l2 = \sqrt{k_B T / k_{OT}}$, the width of thermal distribution, iii) $l3 = v_0 / D_r$ persistence length, distance traveled by characteristic time before the particle rotates. Define a critical active swim speed $v_{0c}$, as a function of trap stiffness/spring constant, drag, and ambient temperature $v_{0c} = \sqrt{k_{OT} k_B T}/\eta$, by $l1 = l2$. By this definition, although effective temperature is much higher than ambient temperature, the contribution of thermal fluctuation can not be ignored at above $v_{0c}$.

The ratio of $l1/l2$ determines whether the system is thermal or active dominated, with $l1 < l2$ being thermal dominated, exhibiting a near Gaussian thermal spatial distribution. In the case of $l1 > l2$, the system is active motion dominated, we then need to examine the ratio of $l1/l3$, which determines whether the active motion is thermalized before rotational characteristic length: $l1 > l3$ indicating the system exhibit a near Gaussian spatial distribution, and $l1 < l3$ exhibiting a double peak distribution.

It is to be noted that the adoption of an optical trap as confinement enabled $l_1$ to be significantly smaller than previous work, yet the large dynamic range of ICEP ($v_0$) allowed us to have and allowed us to investigate in a large range of $l_1$, thus a large range of $l1/l2$ and $l1/l3$. When Thermal fluctuation is comparable to active fluctuation (this work), $l1 \sim l2$, the histogram needs to consider both the contribution from thermal fluctuation and active fluctuation. The result is best treated with the Convolution method, as shown in fig 4a. When thermal fluctuation is much smaller than active fluctuation, $l1 > l2$, the histogram is dominated by active fluctuation. Frydel et. al. theory [16] and Takatory et. al. acoustic trap [17] experiment is in this condition, as shown in Fig 4B. When Thermal fluctuation is larger than active fluctuation, $l1 < l2$, the histogram is dominate by thermal fluctuation, Volpe et. al.'s work [18] is in this condition, as shown in fig 4C.

We defined a critical active swim speed, as a function of trap stiffness/spring constant, drag, and ambient temperature $v0c = \sqrt{k_{OT}k_BT}/\xi$. At speed above this critical speed, the system is dominated by active motion, and at speed lower than this critical speed, the system is dominated by thermal characteristic. By this definition, although effective temperature is much higher than ambient temperature (100 times higher), the contribution of thermal fluctuation can not be ignored at above $v_{0c}$.

### 3.3 Using PSD to Characterize a Single Particle in Confinement

To further study the fluctuation and compare with the existing experiment in biological system characterization [17], we also employed dissipation power spectral density (PSD) analysis to our model system. The PSD reveal the fluctuation from the temporal domain. The fluctuation power spectral density (PSD) is the Fourier transform of the autocorrelation of the particle position, [32]

$$S_x = \frac{1}{T}\int_0^T \int_0^T <x(t)x(t')> e^{i\omega(t-t')}dtdt' \tag{8}$$

where $S_x$ is the power spectral density, T is the total time of integration, $<x(t)x(t')>$ is the auto-correlation function of the particle position $x(t)$ at time t. The PSD represents the intensity of fluctuation at a certain frequency. PSD can be calculated from experimental particle tracking data.

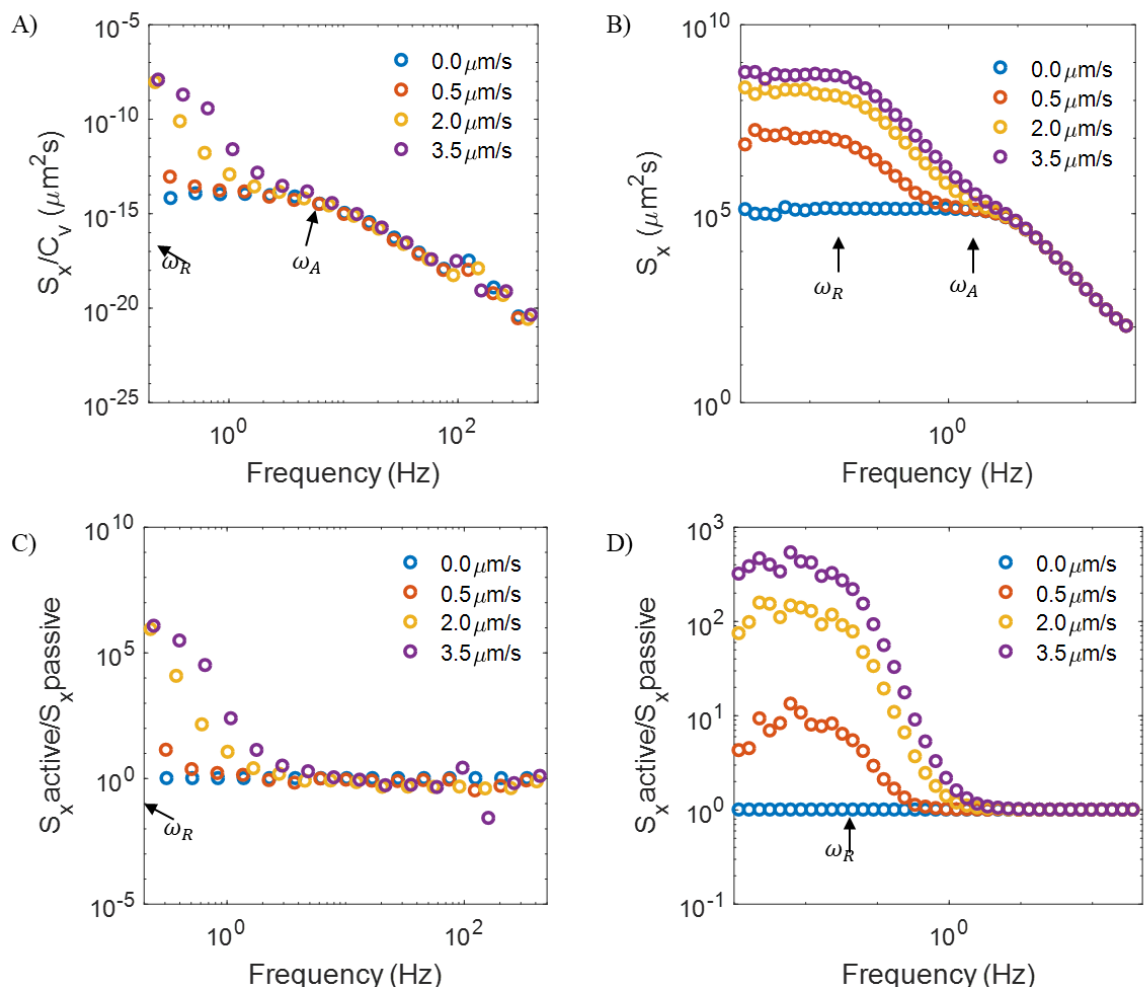

FIG. 5. A) Experimental fluctuation power spectral density (PSD) of 3um Janus particles under 1.5pN/um trap for various AC electric potential at 5 kHz. B) Simulated fluctuation power spectral density (PSD) of an active particle with rotational diffusivity equal to 0.05/s in a quadratic potential with spring constant of 1.5pN/um for various particle speed from 0.0um/s to 3.5um/s. There appears a characteristic frequency as the red line show, determined by the intensity decay to half of the plateau value. C) The effective temperature of 3um Janus particles calculated from experimental data sown in A), as a function of frequency with various AC electric field at 5 kHz. D) The effective temperature of Active particle calculated from simulated data in B), with rotational diffusivity equal to 0.05/s in a quadratic potential with spring constant of 1.5pN/um for various particle speed from 0.0um/s to 3.5um/s.

Fig 5A shows the experimental PSD of the Janus particles in an optical trap. The Brownian particle PSD (the blue circles) is from Jans particle in the trap with zero ICEP voltage, and it exhibits two linear components: a flat plateau at low frequency (0.1~10hz) and a straight line with slope equal to -2 at high frequency (10hz~100hz). The characteristic frequency at which the slope changes corresponds to the competition between thermal fluctuation and optical trap confinement, this is Brownian translational movement and trap force coupling. For an active particle in the same trap, the PSD data (yellow, orange and purple circles) measured from increased ICEP voltage exhibit the same feature as a Brownian particle at higher frequencies (10~100Hz), but at lower frequencies (0.1Hz-10Hz) the PSD value increases with higher active speed. The increase at low frequency can be as high as $10^5$ for an active particle with 8V of applied voltage (3.5 um/s active speed) compared to a Brownian particle.

Fig 5B is PSD from numerical simulation using a similar parameter as experiment and equation (??). It can be seen that the experiment and simulation agree quantitatively at high frequency but only qualitatively at low frequency. In research where PSD was measured, it has always been used qualitatively and not quantitatively: the quality of PSD was investigated rather than quantitative values, and the qualitative observation of a higher power at low-frequency range has been widely adopted to signify active agent vs Brownian particles. [2,6] PSD measurement is highly sensitive to drift noise and at the low frequency (long time scale), the slow drift bias may affect the experimental precision [reference needed]. It is also possible that at long time scale the integration time of experimental measurement is not long enough, leading to noise and inaccuracy at very low frequency[reference needed]. In this work, the quantitative agreement at high frequency and the qualitative agreement at low frequency validate our simulation to explore conditions hard to achieve accuracy in the experiment.

Berg-Sørensen et al theoretically analyzed the PSD of a Brownian particle in a trap, and derived the power spectral density $S(\omega)$ to be a function of $\eta$ the viscosity, $R$ the particle radius, $k_B$ the Boltzmann constant, $k_{OT}$ the trap spring constant, and temperature T. [33,34]

$$S_{Brownian}(\omega) = \frac{12\pi\eta R k_B}{\left(\frac{k_{OT}}{R}\right)^2 + (\eta\omega)^2} T \tag{9}$$

This equation fits well with our PSD data at zero ICEP voltage. when $\frac{k_{OT}}{R} = \eta\omega$, the $S(\omega)$ reach the characteristic frequency at high frequency $\omega_a = k_{OT}/R\eta$.

If we normalize the active particle PSD with Brownian particle PSD, ie., taking the ratio of active PSD and Brownian PSD, we obtain power spectral density ratio $Sx\_ratio(\omega)$, as shown in Fig 5C for experimental data and Fig 5D for simulation data. We found that it has a frequency dependence similar to that of PSD, with a flat plateau at low frequency (0.001~0.1hz) and a straight line with a slope of -2. This reveals a diffusion pattern from the active speed being suppressed in a strong trap at a position where the propelling force is balanced by the trapping force. The high-frequency overlap between active and Brownian particle lead us to keep the high-frequency $Sx\_ratio(\omega)$ at 1. The ratio at zero frequency is always equal to the diffusivity ratio in free space, thus $v_0^2 \tau_r / 2D_t$ should be corresponding to the magnitude of the ratio at low frequency. We can then write the in the following format:

$$S_{Active\_Brownian\_ratio}(\omega) = \frac{v_0^2 \tau_r / 2D_t}{(\omega/\omega_r)^2 + 1} + 1 \tag{10}$$

Where $v_0$ is active speed, $D_t$ is translational diffusivity, and $\tau_r$ is rotational diffusion characteristic time. $\tau_R = 1/D_r$, $\omega_r = 2\pi D_r$. The ratio is shown in Fig 5C for experimental data and Fig 5D for numerical simulation data. Szamel analytically calculated for only active motion [35], and arrived at a format of $S\_ratio(\omega) = \frac{D_f}{C(\omega^2 + \omega_r^2)}$, which is very similar to our results. Unable to include thermal diffusion contribution in his approach, this result would only be able to approximate high active speed, thus high effective temperature limit. With the experimental and simulation data in this work, we can see that PSD of active particles have a Brownian thermal diffusion baseline in addition to the low frequency range multiplier indentified in this work and Szamel's work. [35]

$$S_{Active}(\omega) = S_{Brownian}(\omega) S_{Active\_Brownian\_ratio}(\omega) \tag{11}$$

The data did reveal 2 characteristic frequencies. From a particle movement point of view, the high-frequency feature is determined by translational thermal diffusion and trap coupling, and the low-frequency characteristics are related to Janus particle rotational diffusion and particle active motion. The data from experiments and numerical simulations both show a characteristic frequency $\omega_a$ related to the thermal translational diffusion in the high-frequency range. The low-frequency simulation data show clearly a characteristic frequency $\omega_r$ correspond to the rotational diffusion characteristic time scale $\tau_r$ in MSD plot at long time scale. The experimental data also qualitatively show this feature. The higher active speed the higher rise at low frequency in PSD value compared to Brownian motion, and also larger distance for the active particle to travel away from the center of the trap. It is to be noted that $\omega_r$ is constant across different active speed values.

### 3.4 Effective Temperature: Connecting PSD and HPP

The method of the histogram of particle position (HPP) and fluctuation power spectral density (PSD) analysis have both been employed by scientist to study active system, including artificial active particles [17], and biological system. [2,6,18] HPP gives information on particle spatial distribution and its fluctuations, while PSD gives temporal distribution and fluctuation, and the two yields information of the system that's orthogonal to each other. [35] HPP analysis and PSD analysis, however, were rarely used in the same study together. This is probably because of limitations in experimental measurement and specific feature of the system. For example, in some biological system, the active component (ie., active speed) is small, [6] making the measurement of histogram difficult to differentiate from thermal fluctuation, so HPP measurement is hard and people tend to rely on PSD analysis. On the other hand, in artificial active system such as Janus particles in a trap, the low time resolution makes it hard to obtain large range of PSD data. [17]

It is tempting to use classical thermodynamic concepts to describe an active system, for example, temperature and pressure. The active swim speed would promote people to use an effective temperature to define the motion of the active system, and then use such definition to further study the system characteristics such as viscoelasticity [1,2] and energy dissipation. [36] The concept of effective temperature should in principle signify how active the system is, and lead to a reasonable further analysis of the system properties.

One can define an effective temperature from long-time diffusion in free space, and this is the golden standard of temperature definition for diffusion related process. Fig 6A curve b) is from MSD in free space.

An effective temperature for a confined particle can be defined from HPP:

$$\frac{T_{eff}}{T_0} = \frac{<x^2_{active}>}{<x^2_{passive}>} \tag{12}$$

The integration of Eq 12 can yield an effective temperature that is position independent, or in other words, averaged over the space of the confinement:

$$\frac{T_{eff}}{T_0} = \frac{\int \frac{1}{2}k_{OT}x^2_{active}p(x)dx}{\int \frac{1}{2}k_{OT}x^2_{passive}p(x)dx} \tag{13}$$

One can also define an effective temperature from PSD for a confined particle:

$$\int_0^\infty S_x(\omega)d\omega = \int_0^\infty \frac{1}{T}\int_0^T\int_0^T <x(t)x(t')> e^{i\omega(t-t')}dtdt'd\omega \tag{14}$$

It is to be noted the following equivalence relations: i) Zero frequency PSD = free space thermal diffusion, ii) high-frequency PSD = ambient T, iii) Integral of histogram= integral of PSD.

Mathematically,

$$\int_0^\infty S_{xx} d\omega$$
$$= \int_0^\infty \frac{1}{T}\int_0^T \int_0^T <x(t)x(t')> e^{i\omega(t-t')} dt dt' d\omega$$
$$= \frac{1}{T}\int_0^T \int_0^T <x(t)x(t')> \delta(t-t') dt dt'$$
$$= \frac{1}{T}\int_0^T <x^2> dt \tag{15}$$

The average energy from spatial fluctuation and temporal fluctuation give the same average energy. As shown in Fig 6A curve a) and curve c) overlaps each other, although calculated from Eq 13 and Eq 14 separately. When taking a close look at the HPP or PSD data, one would find that the lost information during the integration process. It is, in general, true that one can not use the integrated form of effective temperature definition, and only in particular conditions can these effective temperature definitions approximate active particle motion, when the effective temperature is used for system properties related to diffusion, such as viscoelasticity, and rubber stiffness.

In the case of the Janus particle in optical trap as in this work, the direction of active ballistic motion is directly linked to rotational diffusivity, and the persistence length being greater than trap thermal length lead to a double-peak distribution of the HPP, so the effective temperature from integration of such HPP would fail as a result. The zero-frequency limit of PSD, as plotted in Fig 6A curve d) is equivalent to long term diffusion in free space as Fig 6A curve b. Experimentally it is hard to measure. One might wonder when to stop measuring and know that the frequency is low enough? Theoretically, the low-frequency PSD should exhibit a plateau as shown in Fig 5B. For an active system, the Brownian component is the same across all active speed, as shown in Fig 6A curve e. The ballistic motion does not affect high-frequency noise and only appear at low frequency, making PSD a more sensitive method to detect active noise than HPP.

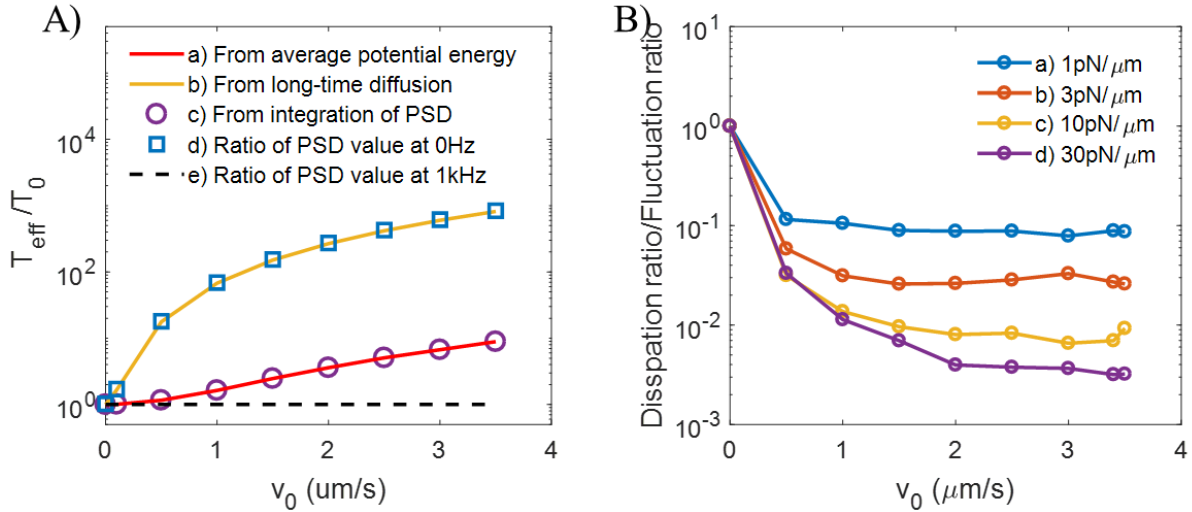

Fig 6. A) Effective temperature $T_{eff}$, normalized with ambient temperature, calculated from simulation data with different definitions: a) $T_{eff}$ calculated from potential energy associated with spatial distribution; b) $T_{eff}$ calculated from diffusion coefficient at long time scale; c) $T_{eff}$ calculated from integrating the PSD curve overall spectral range; d) $T_{eff}$ calculated by evaluating the ratio of active particle PSD over that of Brownian particle at zero frequency (long time scale limit); e) $T_{eff}$ calculated by evaluating the ratio of active particle PSD over that of Brownian particle at 1kHz frequency (short time scale, thermal diffusion dominated);. B) The ratio between energy dissipation and thermal fluctuation, at varying trap stiffness.

PSD gives information on the spectral characteristics of fluctuation-dissipation of energy. [6] When the PSD of a passive particle is well-characterized as a reference, it is possible to compare it with the active particle PSD and get insight in the passive portion and active portion of energy dissipation. From data shown in Fig 5, one can see that for an active particle, the raise in PSD is only at low frequencies, and not in the entire frequency range. We can then conclude that dissipation for an active particle is different from the passive particle. If we take a ratio between active and passive PSD, we will get Fig 6B, the ratio of dissipation over fluctuation. It can be seen that for a Brownian particle which is completely passive, the fluctuation is the same as dissipation, but for an active particle, dissipation is a fraction of fluctuation, shown as the ratio decreases to be smaller than 1, and the higher the active speed, the lower this fraction. While within the same trap a higher active speed result in lower dissipation fraction and this ratio eventually reaches a plateau, a stronger trap will result in a smaller plateau value. The dependence on trap stiffness suggests that the integral of PSD is not the dissipation without a trap. It is possible that the energy may be dissipated through flow in the liquid media around the particle and not being picked up by PSD of the particle. [37]

## 4. Conclusions:

We developed an experimental model system for an active particle confining it in an optical trap. Our system has a large dynamic range by using induced charge electrophoresis of metallic dielectric Janus particle.

Histogram of particle position (HPP) is used to analyze fluctuation in spatial distribution. For a single active particle in an optical trap, we observe the classical Gaussian spatial distribution typical of a Brownian particle when the particle is not driven, and a double-peak distribution when the particle is driven by ICEP electric field and becomes active. We discover that thermal fluctuation cannot be ignored in general, but rather should be incorporated by convolution method, which explains our experimental data as well as prior work. At certain limits, the system is approximated and simplified as a near Gaussian system, and these limits are defined by a critical active speed $v_{0c}=\sqrt{k_{OT}k_BT}/\eta$ and characteristic length scales $l1, l2, l3$, and their ratio.

Fluctuation power spectral density (PSD) is used to analyze fluctuation in temporal distribution. A raise appear at lower frequency and a characteristic time scale $\tau_r$, appear at an extra transition point on PSD, which corresponds to the Janus particle rotational diffusivity. Long-time diffusivities are obtained from the slope of MSD vs. time at the long-time limit without confinement is the same value from the PSD at zero frequency limit.

HPP and PSD methods yield the total energy inside fluctuation are the same. We demonstrate this by integral the total potential energy from HPP by Eq 13 and total energy from PSD by Eq 14. Also, the integration of PSD, which used to be considered as the dissipation of energy, is dependent on the trap stiffness. Thus, the use of this "dissipation of energy" need to be more careful in active system.

Unlike a living cell, this model system has a well-defined linear force trap with a known active particle. We distinguish the fluctuation being purely from the particle and demonstrate the breakdown of fluctuation-dissipation relation. Since active matter plays a central role in many systems, our findings show limitations to apply thermodynamics in their classical formulation to study systems with active, non-thermal noise.